\documentclass[12pt, a4paper]{article}

\usepackage{amsmath}
\usepackage{amsfonts}
\usepackage{amssymb}
\usepackage{graphicx, rotating}
\usepackage{epstopdf}
\usepackage{epsfig}
\usepackage{latexsym}
\usepackage{graphicx}
\usepackage{color}
\usepackage{amsmath,bm,amssymb}
\usepackage{cite}
\usepackage{slashed}
\usepackage{hyperref}
\definecolor{rossos}{cmyk}{0,1,1,0.55}
\definecolor{bluscuro}{rgb}{0.15, 0.2, .85}
\definecolor{bluchiaro}{cmyk}{1,.3,0.,0.1}
\hypersetup{colorlinks, citecolor=bluscuro, linkcolor=black, urlcolor=bluscuro}

\newcommand{\be}{\begin{equation}}
\newcommand{\ee}{\end{equation}}
\newcommand{\bea}{\begin{eqnarray}}
\newcommand{\eea}{\end{eqnarray}}

\def\pt{\partial}

\begin{document}

\begin{titlepage}
\begin{flushright}
IFT-UAM/CSIC-20-86
\end{flushright}
\vspace{.3in}

\vspace{1cm}
\begin{center}
{\Large\bf\color{black} Proof of a Momentum/Complexity Correspondence}\\

\bigskip\color{black}
\vspace{1cm}{
{\large J.~L.~F. Barb\'on, J. Mart\'{\i}n-Garc\'{\i}a  and M.~Sasieta}
\vspace{0.3cm}
} \\[7mm]
{\it {Instituto de F\'{i}sica Te\'orica IFT-UAM/CSIC,\\ c/ Nicol\'as Cabrera 13, Universidad Aut\'onoma de Madrid, 28049, Madrid, Spain}}\\

{\it E-mail:} \href{mailto:jose.barbon@csic.es}{\nolinkurl{jose.barbon@csic.es}}, \href{mailto:javier.martingarcia1@gmail.com}{\nolinkurl{javier.martingarcia1@gmail.com}},  \href{mailto:martin.sasieta@csic.es}{\nolinkurl{martin.sasieta@csic.es}}\\
\end{center}

\bigskip

\vspace{.4cm}

\begin{abstract}
We show that the holographic Complexity $=$ Volume proposal satisfies a very general notion of 
Momentum/Complexity correspondence (PC), based on the Momentum Constraint of General Relativity. It relates  the rate of complexity variation with an appropriate matter momentum flux through spacelike extremal surfaces. This formalizes the intuitive idea that `gravitational  clumpling' of matter increases
complexity, and the required notion of `infall momentum'  is shown to have a  Newtonian avatar which expresses this idea.  The proposed  form of the PC correspondence  is found to be  exact  for  any solution of Einstein's equations in $2+1$ dimensions, and any spherically symmetric solution in arbitrary dimensions, 
generalizing all previous calculations using spherical thin shells.  
Gravitational radiation enters through a correction
which does not have a straightforward interpretation as a PC correspondence. Other obstructions to an exact PC duality have a topological origin and arise in the presence of wormholes. 
\end{abstract}
\bigskip

\end{titlepage}


\section{Introduction} 

\noindent

Quantum complexity has been identified as a key notion in the development of the holographic dictionary  for its promise to
offer a peek into the interior of black holes  \cite{suss1}. In the so-called Complexity $=$ Volume (VC) prescription, the complexity of a state with given  boundary data is to be computed in the bulk as the volume of the extremal spacelike hypersurface anchored at those  boundary data \cite{VC}. The basic motivation for this proposal is the observation that it measures the growing volume of black-hole interiors, in direct analogy with the growth of complexity of tensor network representations of many-body quantum states with thermal properties \cite{harmal, epr=er}. 

A recurring theme in this context  is the notion that gravitational `clumping' increases complexity of the dual quantum state. If a black hole is formed, this is
realized in the most extreme way, as the complexity keeps growing linearly well after the black hole has equilibrated its exterior geometry. However, the growth of complexity  occurs  for any gravitational infall of matter, however dilute, as indicated by explicit calculations for collapsing thin shells. A time-reversal transformation to a situation with matter outflow should instead decrease the complexity, suggesting that there is a relation between some average  `infall momentum' and the rate of complexity change. 

Explicit Momentum/Complexity  (PC) relations for states or operators have been described for low-dimensional models \cite{susn, javi, linmalzhao, mousatov}  and thin-shell approximations in higher-dimensional models (cf. for example \cite{thinsh, us, susskind5}). In all these situations, the relevant radial momentum arises as some canonical momentum in an effective $1+1$ dimensional effective Lagrangian, and the detailed form of the PC correspondence depends to some extent on the coordinates chosen and the dynamical assumptions on the nature of the shells (massive dust, null dust, branes, etc). It was pointed out in \cite{us} that, using the VC extremal surfaces themselves as a time foliation of the bulk,  it is possible to write any PC relation of thin shells in the form 
\be\label{shell PvC}
{\dot{\cal C}}_{\rm shells} = - \int_{\Sigma_t} N^\mu_\Sigma \, T_{\mu\nu}\,  C^\nu_\Sigma \;,
\ee
where $\Sigma_t$ is the extremal-volume hypersurface at time $t$, the dot indicates time derivative with respect to this asymptotic time variable, $N_\Sigma$ is the unit timelike normal and  $C_\Sigma$ is an inward pointing radial field, tangent to $\Sigma_t$, whose modulus at each point is the radius of the angular sphere: $|C_\Sigma | = r$. We will refer to this form of the PC correspondence as the PVC relation, because of the prominent role played by the maximal-volume surfaces  in its definition. A key property of (\ref{shell PvC}) is that all dynamical assumptions about the shells are concealed inside the energy-momentum tensor $T_{\mu\nu}$. Therefore, it is natural to suspect that a PVC relation of this form could have a much wider degree of generality. In this paper we confirm this expectation, showing that the content of (\ref{shell PvC}) is essentially the Momentum Constraint of General Relativity (GR).

\section{PVC from the Momentum Constraint}

\noindent

We shall work  with spacetimes $X$ asymptotic to global AdS$_{d+1}$ with $d\geq 2$.  The bulk state is described as a solution of Einstein equations with energy momentum tensor $T_{\mu\nu}$, and the asymptotic behavior of a normalizable state. 
We shall adopt units such that the asymptotic radius of curvature of AdS is $\ell_{\rm AdS} =1$, although most of our results still hold in the flat spacetime limit $\ell_{\rm AdS} \rightarrow\infty$. The   VC formula is taken to be
\be\label{vc}
{\cal C}\,[\Sigma_t] = {d-1 \over 8\pi G} \,{\rm Vol} (\Sigma_t)\;,
\ee
a regularized volume of an extremal codimension-one hypersurface $\Sigma_t$, anchored at boundary time $t$, which labels the real line in ${\bf R} \times {\bf S}^{d-1}$, the conformal boundary of $X$. For notational simplicity we will often suppress the time label in $\Sigma_t$, with the implicit understanding that a choice of $\Sigma$ is equivalent to a choice of boundary time. 

To fix notation,  $g_{\mu\nu}$ denotes  the metric on $X$ and  $h_{ab}$ the induced metric on $\Sigma$, with world-volume coordinates $y^a$. Latin indices are raised and lowered with $h_{ab}$, whereas greek indices are operated with $g_{\mu\nu}$.  The   embedding of $\Sigma$ into $X$ is described by  the functions $X^\mu (y^a)$, with tangent frame vector fields $e^\mu_a = \partial_a X^\mu$. The extrinsic curvature of $\Sigma$ is denoted $K_{ab}$, and its trace  $K = h^{ab} K_{ab}$ will vanish throughout our discussion, since we are focusing on extremal-volume surfaces.  Finally, the future-pointing, unit timelike normal to $\Sigma$ is denoted
$N^\mu_\Sigma$.  

We begin by deriving a useful equation for the rate of VC.  Since $\Sigma$ is extremal, the first-order variation of (\ref{vc}) with respect to the embedding fields $X^\mu (y^a)$ only picks a boundary term depending on the  variation of the anchoring surface: 
\be\label{bt}
\delta {\cal C}\,[\Sigma] = {d-1 \over 8\pi G} \int_{\partial \Sigma} \delta X_\Sigma\;,
\ee
where $(\delta X_\Sigma)_a = e^\mu_a \,\delta X_\mu$ is the embedding variation, pulled back to $\Sigma$. For a rigid time translation $\delta t$  at the boundary,
we have $\delta X_\Sigma = \delta t \,(\pt_t)_\Sigma$, where $\partial_t$ denotes the time-translation vector in $X$, which is asymptotically a Killing vector. Dividing by $\delta t$ we obtain an ADM-like expression for the rate of VC:
\be\label{ratevc}
{\dot{\cal C}}\,[\Sigma] = {d-1 \over 8 \pi G} \int_{\pt \Sigma} (\pt_t)_n = {d-1 \over 8 \pi G} \int_{\pt \Sigma}  dS^a \, e^\mu_a \,(\pt_t)_\mu \;.
\ee
This equation represents the complexity rate as the integral of $(\pt_t)_n = e_n \cdot \pt_t $ over the boundary of the extremal surface, where $e^\mu _n = e^\mu_a \,n^a_{\pt \Sigma}$, with $n_{\pt \Sigma}$ the outward pointing normal to $\pt \Sigma$. Since $e_n$ is tangent to $\Sigma$, the integrand is sensitive to the asymptotic bending of $\Sigma$ by the presence of non-trivial geometry in the bulk. More precisely, we pick the term of order $1/r^{d-1}$, for $r$ the radius of an angular sphere which regularizes $\pt \Sigma$. 

Given any `current' $J^a$ defined on $\Sigma$, which has the same boundary integral as $(\pt_t)_n$, 
\be\label{samei}
\int_{\pt \Sigma} J_n = \int_{\pt \Sigma} (\pt_t)_n\;,
\ee
we can use Stokes theorem to write the VC rate  as a bulk integral of its `source' over the extremal surface:
\be\label{bu}
{\dot{\cal C}}\,[\Sigma] = {d-1\over 8\pi G} \int_{\Sigma} \nabla_a J^a\;.
\ee
A strategy to obtain a PVC equation is to make a clever choice of $J$, in such  way that  it is sourced by a momentum density. A simple example is provided by the well-known case of spherical thin shells, whose PVC relation (\ref{shell PvC}) can  be derived in this language by choosing $J_a = (\pt_t)_\mu \,e^\mu_a $. In this approximation scheme  $\pt_t$ is a Killing vector except for jumps at the worldvolume of the shells, so that the integral (\ref{bu}) localizes to delta-function contributions, with coefficients controlled by the  junction conditions (cf. \cite{poisson} for a review). This derivation shows that the PVC relation is independent of any choice of equation of state on the world-volume of the shells  (cf. \cite{us}). 

\subsubsection*{Exact PVC}

\noindent

In order to pursue this strategy in more general terms, we
can work backwards by seeking a natural GR equation that uses the momentum density over a spacelike surface. The obvious candidate is  the so-called Momentum Constraint (MC): given any Cauchy surface $\Sigma$, initial data $h_{ab}$ and $K_{ab}$ are constrained by the equation (cf. \cite{poisson})
\be\label{mc}
\nabla^a K_{ab} - \nabla_b K = -8\pi G \,{\cal P}_b\;,
\ee
where ${\cal P}_b = -N^\mu_\Sigma\,T_{\mu\nu} \, e^\nu_b $ is the pulled-back momentum flux through $\Sigma$. For the purposes of this work, we can simplify this equation by setting $K=0$, since  $\Sigma$ is taken to be extremal. 

\begin{figure}[t]
	\centering
	\includegraphics[width=8cm]{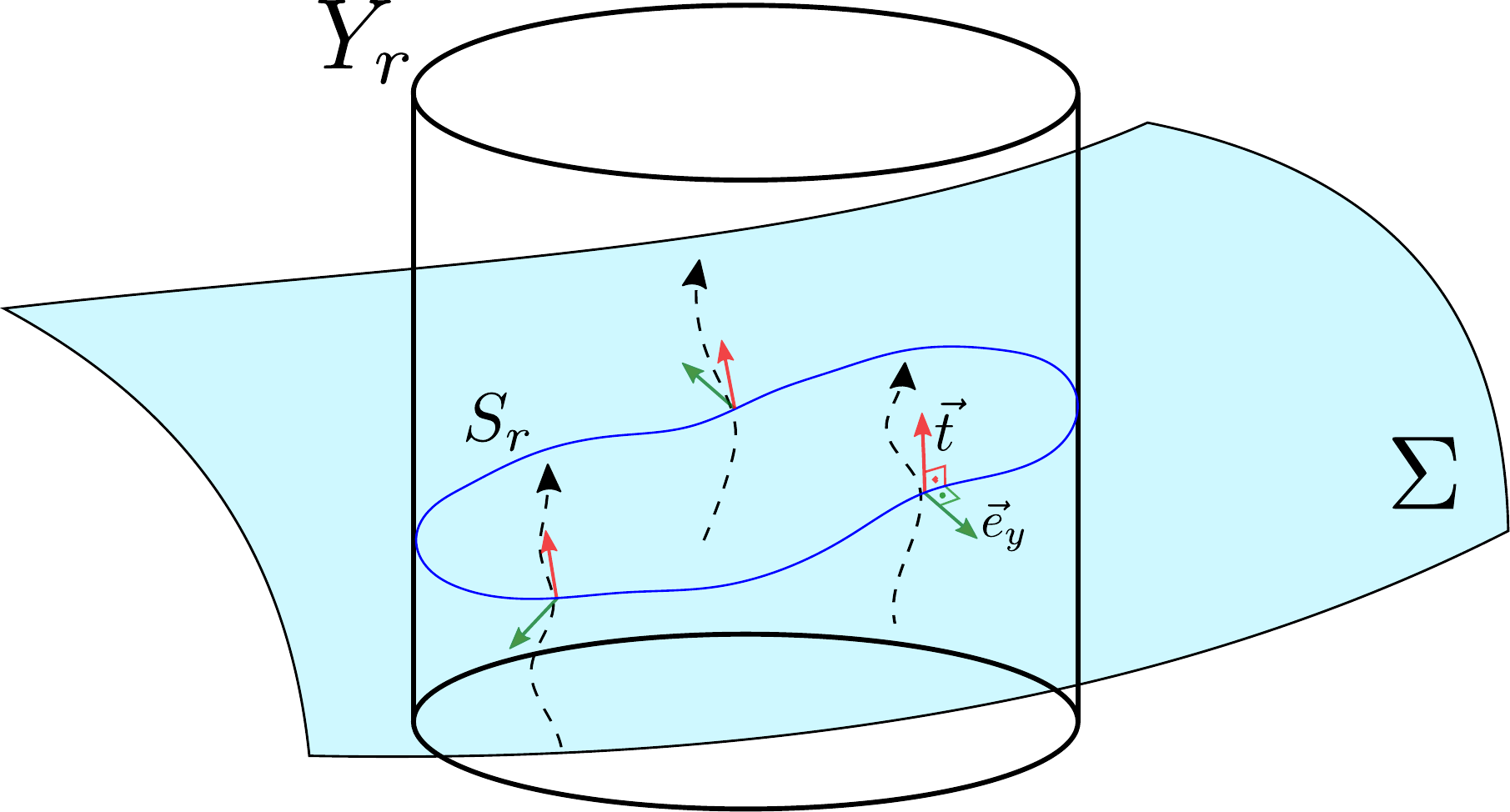} 
	\caption{\emph{For each point in $S_r$, the time coordinate is chosen to properly parametrize a geodesic (dashed lines) on $Y_r$. $\vec{e}_y$ is then picked to be orthogonal to $S_r$.}\label{fig:cilindrito}}
\end{figure}

In order to integrate the MC we must introduce a tangent vector field on $\Sigma$. Anticipating its role in what follows, we shall refer to this field,  $C_\Sigma$, as the `infall' vector field, despite the fact that at this point it is completely arbitrary.  Multiplying (\ref{mc}) by $C^b$ and integrating by parts we obtain the equivalent expression 
\be\label{mcin}
\int_\Sigma {\cal P}_C = -{1\over 8\pi G} \int_{\pt \Sigma } dS^a \,K_{ab}\, C^b  +{1\over 8\pi G} \int_{\Sigma} K^{ab} \,\nabla_a C_b\;,
\ee
where ${\cal P}_C = {\cal P}_a \, C^a$ is the  momentum component that is being selected by the $C$-field. 
The left hand side has the form of the momentum integral we are seeking, whereas we have a boundary term in the right-hand side that we could try to interpret as ${\dot{\cal C}}$. In other words, we would like to set 
\be\label{jota}
J_a = -{1 \over d-1} \,K_{ab}\,C^b\;,
\ee
and fix the behavior of $C^b$ at the boundary so that we satisfy (\ref{samei}). This can be analyzed by means of a local computation as follows. In the vicinity of $\Sigma$, we may choose coordinates such that the metric reads 
\be\label{fg}
ds^2_X \rightarrow {dr^2 \over r^2} + r^2 \left(-dt^2 + \gamma_{ij} (r, t, \theta) \,d\theta^i \,d\theta^j \right) \;{\rm as}\;\;r\rightarrow \infty\;.
\ee
Here,  $r$ is a Fefferman--Graham coordinate which foliates $X$ by timelike codimension-one submanifolds $Y_r$. The angles $\theta^j$ parametrize the intersection $S_r = Y_r \cap \Sigma$, of spherical topology and induced metric proportional to $\gamma_{ij}$, which is itself asymptotic to a unit round $(d-1)$-sphere,  up to normalizable corrections of order   $1/r^d$. The crucial simplifying property  of (\ref{fg})  is the choice of time coordinate, which is geodesic and orthogonal to $S_r$ (cf. Figure \ref{fig:cilindrito}). 

The   induced metric on $\Sigma$ can be written near the boundary as 
\be\label{induced}
ds^2_\Sigma \rightarrow dy^2 + r^2 (y) \,\gamma_{ij} (y, \theta) \,d\theta^i \,d\theta^j \;,
\ee
for some function $r(y)$ asymptotic to $\sinh(y)$ as $y\rightarrow \infty$.
This allows us to write the normal one-form  as 
$N_\Sigma = e^t_y \,dr - e^r_y \,dt$, and compute the extrinsic curvature $K_{ab} = e^\mu_a \,e^\nu_b \,\nabla_\mu \,N_\nu$. The relevant component turns out to be
$K_{yy}$ which, using the traceless character, $K=0$, may be evaluated as $K_{yy} = - r^{-2} \,\gamma^{ij} \,K_{ij}$. Explicitly 
\be\label{ext}
K_{yy} = -(d-1)\,r\,e^t_y - {1\over 2r^2} \,e^r_y \,\gamma^{ij} \pt_t \gamma_{ij} - {r^2 \over 2} \,e^t_y \,\gamma^{ij} \pt_r \gamma_{ij}\;.
\ee
An asymptotic  analysis reveals the large-$r$ scalings  $ e^r_y \sim r$, $ e^t_y \sim 1/r^{d+1}$, $\pt_t \gamma_{ij} \sim 1/r^d$ and $\pt_r \gamma_{ij} \sim 1/r^{d+1} $, so that the right hand side of (\ref{ext}) is dominated by the first term: $K_{yy} \approx -(d-1) \,r\,e^t_y$.  Since $e_y \cdot \pt_t = -r^2 \,e^t_y\,$, we learn that (\ref{samei}) can be satisfied provided the $C$-field is chosen with the boundary conditions
\be\label{boc}
C_{\Sigma} \rightarrow -r(y) \,\pt_y\; \;\;\; {\rm as}\;\;\;\;y\rightarrow \infty\;. 
\ee
This is exactly the result that was found `empirically' for the case of thin shells in \cite{us}, justifying  the name `infall field' which, from this point of view, is nothing but the condition for the integrated Momentum Constraint to compute the complexity rate.

We are now ready to assemble all the pieces and write down a `generalized PVC' relation. Defining a total `$C$-momentum' through $\Sigma$ and a `remainder' by the expressions 
\be\label{pc}
P_C \,[\Sigma] = \int_\Sigma {\cal P}_C \;, \qquad 
R_C \,[\Sigma] = -{1\over 8\pi G} \int_\Sigma K^{ab} \,\nabla_a C_b\;,
\ee
we have established 
\be\label{genPVC}
{\dot {\cal C}}\,[\Sigma] = P_C \,[\Sigma] + R_{C} \,[\Sigma]\;.
\ee
This shows that part of the complexity rate at time $t$ can always be attributed to momentum flow through $\Sigma_t$.  In fact, a sufficient condition can be placed on the `infall field'  which ensures the vanishing of the remainder. The extrinsic curvature $K_{ab}$ being symmetric and traceless, we can write the remainder in the form 
\be\label{remcl}
R_C \,[\Sigma] = -{1\over 8\pi G} \int_{\Sigma} K^{ab} \; \left(\nabla_{(a} C_{b)}  - {1 \over d} \, h_{ab} \, \nabla \cdot C \right) \;.
\ee
The term in parenthesis is proportional to the conformal Lie derivative, which vanishes  if the $C$-field is a  conformal Killing vector (CKV). This happens for any smooth spherically symmetric state, for which the induced metric on $\Sigma$ can be written as $ds^2_\Sigma = dy^2 + r(y)^2 \,d\Omega_{d-1}$ for some function $r(y)$. In this case the field $C_\Sigma = -r(y) \partial_y$ is an exact CKV   throughout $\Sigma$.  The same is true of any solution of Einstein's equations in $2+1$ dimensions, because $\Sigma$ is then two-dimensional. In both these cases, when $\Sigma$ has the topology of a $d$-dimensional ball, the induced metric on $\Sigma$ is conformal to the Poincar\'e ball $ds^2_{\rm ball} = dz^2 + \sinh^2 (z) \,d\Omega_{d-1}^2$, with a rescaling factor which approaches unity at $\pt\Sigma$. The  Poincar\'e ball provides a `canonical' infall field $C_\Sigma = -\sinh(z) \pt_z$ which is a radial CKV on $\Sigma$ with the appropriate boundary conditions (\ref{boc}). A subtlety occurs when $\Sigma$ has non-trivial topology, such as the wormhole of an eternal black hole, a situation which will be explained in the last subsection below. 

Therefore, we conclude that any spacetime in $2+1$ dimensions and any spherically symmetric state in arbitrary dimensions satisfies an exact PVC relation 
${\dot{\cal C}} = P_C\, [\Sigma]$.  It is notable that we obtained all these results with no extra hypothesis on the nature of the matter, i.e. no positivity conditions on $T_{\mu\nu}$ were required. This suggests that the nature of the PVC relation is essentially  kinematical  once we take into account the constraints of GR. 

\subsubsection*{A Newtonian Interpretation}

\noindent

For any state satisfying an `instantaneous' PVC relation, the radial CKV field is conformal to the canonical $C$-field of the Poincar\'e ball, which vanishes at the `center'.  
This vanishing point  may be moved by the action of the asymptotic isometries, such as translations in Minkowski spacetime, but a given globally defined infall field will always have a `center'. This suggests that the infall momentum behaves like angular momentum does: an arbitrary center must be specified, although any center is a valid reference point. 

The important  notion of `infall momentum' can be further elucidated by taking the Newtonian limit for a collection of point particles. We can have  these particles moving deep inside AdS,  in
a region of size $\ell \ll \ell_{\rm AdS}$ or work directly in asymptotically flat spacetime. In the Newtonian approximation we can neglect static or dynamic curvature effects and the associated back reaction of $\Sigma$, which can be taken to be a spacelike slice of flat intrinsic geometry. Fixing the  reference system at the point where $C_\Sigma =0$, the complexity rate in the Newtonian approximation is the total  infall momentum for the particle system:
\be\label{innew}
{\dot{\cal C}}_{\rm Newtonian} =P_{\rm infall}  =  -{1\over \ell_{\rm AdS}} \sum_i {\bf x}_i \cdot {\bf p}_i \;,
\ee
where we have momentarily restored the dependence on the `box' length scale $\ell_{\rm AdS} =1$, an arbitrary choice in this Newtonian discussion. 
We see that it is indeed a sort of `radial-inward' version of the angular momentum, constructed with scalar products rather than vector products. Just like angular momentum, the so-defined `infall momentum'  is not invariant under translations or boosts, and a special role is played by the center of mass ${\bf X} = \sum_i m_i {\bf x}_i / \sum_i m_i$. Suppose our system has a number of distant clusters, so that each of them can be regarded as approximately isolated.   The total infall momentum can be decomposed in `intrinsic' and `orbital' parts: 
\be\label{cmm}
P_{\rm infall} = \sum_\alpha P_{\rm infall} [{\bf X}_\alpha] - \sum_\alpha {\bf P}_\alpha \cdot {\bf X}_\alpha\;,
\ee
where ${\bf X}_\alpha$ is the center of mass of the $\alpha$-cluster and  ${\bf P}_\alpha$ its total momentum. In this expression, $P_{\rm infall}[{\bf X}_\alpha]$ accounts for the `intrinsic'  infall momenta within each cluster, measured with respect to its center of mass. Hence, `compositeness' of effective particles is incorporated through an additive term for each particle, something analogous to `spin'. 

Unlike angular momentum, infall momentum will not be conserved in general. Its time derivative, proportional to the second derivative of the complexity, is 
\be\label{deri}
{\dot P}_{\rm infall} = -2 T - \sum_i {\bf x}_i \cdot {\bf F}_i \;,
\ee
where ${\bf F}_i$ is the Newtonian force acting on the $i$-th particle and $T$ is the total kinetic energy. If the internal dynamics of the system is described
by a potential $V({\bf x}_i)$ which is a homogeneous function of degree $k$, Euler's theorem implies
$
{\dot P}_{\rm infall} = -2 T +k V=-(k+2) T + kE$, with $E$ the conserved total energy.  For a  gravitational  system, $k=-1$, which is either unbound or marginally bound, $E\geq 0$, the time derivative of the infall momentum, or equivalently,  the second derivative of the complexity, is guaranteed to be negative. This suggests that there could be inequalities for $d^2 {\cal C} /dt^2$ coming from positive energy conditions.

Infall momentum has the crucial property of being a    total derivative,
$P_{\rm infall} = {\dot{\cal I}}_{\rm clump}$, where
\be\label{mi}
{\cal I}_{\rm clump} = -{1\over 2} \sum_i m_i {\bf x}_i^2
\ee
is a sort of `spherical' moment of inertia which measures the degree of `clumping' of the matter. Hence we find that, within   the Newtonian approximation, the complexity is completely determined, up to an additive constant,  by the degree of matter `clumping'. 
\be\label{ncom}
{\cal C}_{\rm Newtonian} = {\cal C}_{0} + {\cal I}_{\rm clump}
\;.
\ee

\subsubsection*{Obstructions}

\noindent

  The most important  exception to an exact PVC relation is provided by gravitational waves. In this case, the Weyl tensor of $X$ does not vanish, and embedded hypersurfaces will in general fail to be conformally trivial. In the absence of a  canonical choice of $C_{\Sigma}$ in the bulk, a remainder correction will be present generically. This is natural from the physical point of view, since a black hole could be formed by colliding gravitational waves, and the  linear growth of complexity must eventually build up at long times even if $T_{\mu\nu} =0$ all along. In such a situation, the rate of complexity change would be  completely given by the remainder term $R_C [\Sigma]$ in (\ref{genPVC}).

  Approximate PVC relations should exist in the context of the linearized gravity approximation. If $X$ contains  gravitational waves perturbing
  a spherically symmetric  background $X_0$, it should be possible to establish an approximate PVC relation of the form 
  \be\label{apppvc}
  {\dot{\cal C}}\,[\Sigma_0] \approx -\int_{\Sigma_0} N^\mu_{0} \; (T_{\mu\nu} + t_{\mu\nu} ) \; C^\nu_{0}\;, 
  \ee
  where $t_{\mu\nu}$ is a pseudotensor of Landau--Lifshitz type and the normal, $N_0$, and infall, $ C_0$, vectors  are referred to the  surface $\Sigma_0$, extremal with respect to the  background geometry $X_0$. If the gravitational waves can be fully related to matter sources, the $t_{\mu\nu}$ contribution will be hierarchically smaller than the matter contribution.
  
  A different type of obstruction to an exact PVC correspondence occurs when we have wormholes. The simplest example which captures the relevant issues is the Einstein--Rosen bridge of an eternal black hole. In vacuum, the extremal surfaces are spherically symmetric cylinders of topology ${\bf R} \times {\bf S}^{d-1}$,  with
  a ${\bf Z}_2$ reflection symmetry between left and right sides, acting on ${\bf R}$ in the standard fashion. Radial CKVs exist, but the asymptotic boundary conditions are necessarily incompatible with the `infall' interpretation in both boundaries: if the $C_\Sigma$ field is `infall' on the right side, it must be `outfall' on the left side. Revisiting  the asymptotic boundary conditions for the $C$-field (\ref{samei}) and (\ref{jota}) we see that an inversion of $C$ is correlated with an inversion of the time-translation vector $\pt_t$, namely the equation ${\dot {\cal C}}\,[\Sigma] = P_C \,[\Sigma]$ holds when
  we interpret the complexity rate as measured with respect to the Killing Hamiltonian $H_K = H_R - H_L$. In this case one obtains ${\dot{\cal C}}_K =0$ for
  the vacuum solution, where the $K$ label stands for the choice of time variable dual to $H_K$.  The same is true for any ${\bf Z}_2$-symmetric momentum configuration, such as identical collapsing matter distributions on both sides. In order to get ${\dot{\cal C}}_K > 0$ we need a sufficient amount of `outfall' in the left side.  
  
  \begin{figure}[h]
  	\centering
  	\includegraphics[width=8cm]{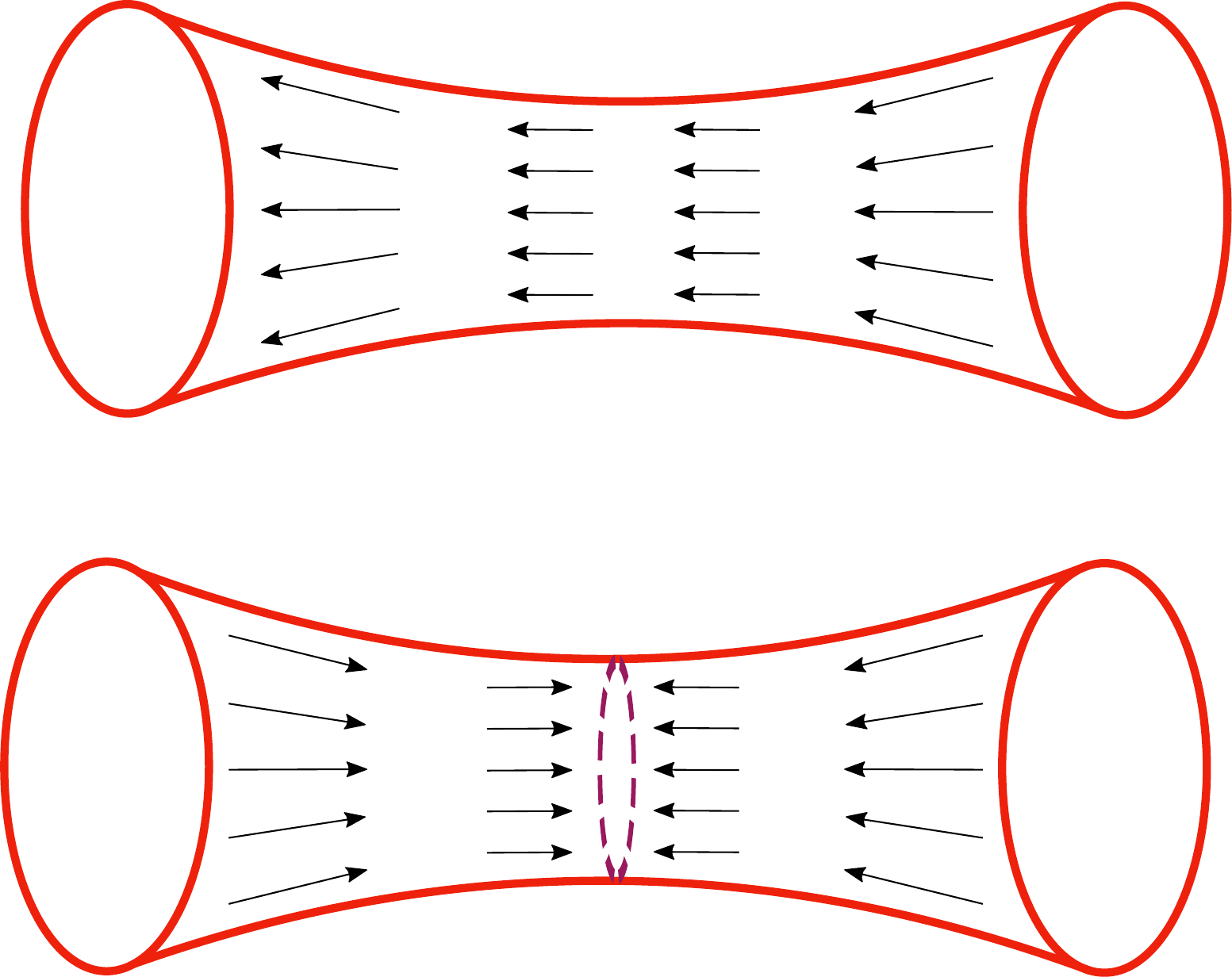} 
  	\caption{\emph{On the top panel, a CKV field on the Einstein--Rosen bridge is in-falling on one side and `out-falling' on the other. On the bottom panel, insisting on being in-falling on both sides forces a discontinuous jump through a defect in the interior.}\label{fig:wh}}
  \end{figure}

  For the case of a vacuum Einstein--Rosen bridge, it is certainly possible to define CKVs  with appropriate infall conditions in the vicinity of each boundary, but these choices are necessarily incompatible with each other in the bulk; at some point the conformal Lie derivative must be non-zero, and a contribution from the remainder is turned on. For instance, if we want to compute the standard complexity rate with respect to the TFD Hamiltonian $H_{\rm TFD} = H_R + H_L$, we must
  introduce a defect in the interior along which the $C$-field switches its orientation from `right-infall' to `left-infall' (cf. Figure \ref{fig:wh}). If we do this at the  ${\bf S}^{d-1} $  sitting at the 
  fixed point of the ${\bf Z}_2$ reflection, we obtain a delta function contribution to the integrand of the remainder. A simple calculation reveals then the
  standard result ${\dot{\cal C}}_{\rm TFD} = 2M$.

\section{Conclusions and Outlook}

\noindent

In this paper we have shown that a Momentum/Complexity correspondence is implicit in the Complexity=Volume prescription, as a result of the Momentum Constraint in GR. This PVC correspondence is based on two ingredients that were advanced in the thin-shell analysis of \cite{us}: the use of maximal-volume hypersurfaces as the time foliation to measure the momentum, and a particular choice of momentum component along the extremal surfaces, determined by an appropriate `infall field' $C_\Sigma$. In formulas
\be\label{PVC}
{\dot {\cal C}}\,[\Sigma] = \int_\Sigma {\cal P}_C + R_C \,[\Sigma]
\;.
\ee
The infall field is required to have fixed boundary conditions at infinity,  but otherwise the freedom implicit in its specification is reflected in the existence of a `remainder' correction $R_C [\Sigma]$ to the PVC relation. The remainder vanishes if $C_\Sigma$ extends to the bulk   as a CKV, something that is guaranteed  for any spacetime in $2+1$ dimensions and any spherically symmetric spacetime  in arbitrary dimensions. From the physical point of view, the most important exception is provided by gravitational waves. The existence of a remainder term which does not admit a simple interpretation in terms of `infall momentum' is actually natural, since we know that there is simply no candidate for a local measure of purely gravitational momentum to be integrated over $\Sigma$. The best one could expect is to define such `local gravitational momentum'  in perturbation theory, as in (\ref{apppvc}). 
An alternative route is to generalize the ideas of this paper as follows: the  Momentum Constraint is the trace of the more general Codazzi equation,
\be\label{gcod}
\nabla_c K_{ab} - \nabla_b K_{ac} = W_{abc} + {8\pi G \over d-1} (h_{ab} {\cal P}_c - h_{ac} {\cal P}_b)\;,
\ee
where $W_{abc} = N^\mu_\Sigma W_{\mu\nu\rho\sigma} \,e^\nu_a e^\rho_b e^\sigma_c$ is the contracted Weyl tensor, pulled back to $\Sigma$. A generalized PVC correspondence follows by integrating the Codazzi equation, rather than the Momentum Constraint, against a three-index `infall tensor' $M^{abc}$. This tensor can be chosen with the same symmetries as the pulled-back Weyl tensor $W_{abc}$ and satisfying a projected covariant constancy condition: $K_{ab} \nabla_c M^{abc} =0$. Then, following the same steps which lead to (\ref{genPVC}), one finds the generalized PVC relation
\be\label{gpvc}
{\dot {\cal C}} = -\int_\Sigma N^\mu_\Sigma \; T_{\mu\nu} \;C^\nu_\Sigma - {d-1 \over 16\pi G} \int_\Sigma W_{abc} \,M^{abc}_\Sigma \;,
\ee
where $C^b = h_{ac} M^{abc}$ is an infall  vector field induced by the `infall tensor field'. In this form, the generalized PVC relation does pick a contribution from  gravitational waves through their Weyl tensor, while at the same time recovering the matter contribution as a local integral over infall momentum. The details of this generalization will be reported in 
\cite{toappear}.

Coming back to the PVC relation presented in this paper, we have shown that the central concept of `infall momentum' has a Newtonian version which explicitly captures the intuitive idea that matter clumping increases complexity. This provides an interesting perspective on the `second law of complexity'  \cite{suss3}. The structure of the PVC relation also suggests that it may be useful in connection with the so-called `first law of complexity' \cite{firstl}.  Finally, it would be interesting to see if an analogous PC correspondence with the same degree of generality exists for the Complexity $=$ Action proposal \cite{AC}.

 \section{Acknowledgments}
\label{ackn}
\noindent

This work is partially supported by the Spanish Research Agency (Agencia Estatal de Investigaci\'on) through the grants IFT Centro de Excelencia Severo Ochoa SEV-2016-0597,  FPA2015-65480-P and PGC2018-095976-B-C21. The work of J.M.G. is funded by 
Fundaci\'on La Caixa under ``La Caixa-Severo Ochoa'' international predoctoral grant. The work of M.S. is funded by the FPU Grant FPU16/00639.

\end{document}